\documentclass[aps,prl,floatfix,twocolumn,superscriptaddress]{revtex4-1}

\usepackage{graphicx,
            caption,
            subcaption,
            amsmath,
            braket,
            units,
            ragged2e,
            xcolor,
            xspace,
            nicefrac,
            lipsum,
            nameref,
            textcomp,
            urwchancal,
            upgreek,
            isotope,
            hyperref,
            cleveref,
            siunitx
            }

\DeclareCaptionLabelSeparator{dot}{. }
\makeatletter
\def\justified{
	\let\\\@normalcr
	\@rightskip\z@skip \rightskip\@rightskip
	\leftskip\z@skip
	\parindent 0em\relax
	\setlength{\parfillskip}{0pt plus 1fil}}
\DeclareCaptionJustification{justified}{\justified}
\hypersetup{colorlinks=true, linkcolor=blue,citecolor=blue,urlcolor=blue}
\def\unit #1 #2 {\SI{#1}{#2}\xspace}
\def\range #1 #2 #3 {\SIrange{#1}{#2}{#3}\xspace}
\DeclareSIUnit\gauss{G}

\newcommand{\myref}[2][]{Fig.~\hyperref[#2]{\ref*{#2}#1}}
\newcommand{\Myref}[2][]{Figure~\hyperref[#2]{\ref*{#2}#1}}
\newcommand{\Mytabref}[2][]{Table~\hyperref[#2]{\ref*{#2}#1}}

\begin{document}

\title{Feshbach Resonances in an Erbium-Dysprosium Dipolar Mixture}

\date{\today}
	
\author{Gianmaria Durastante}
\affiliation{
    Institut f\"{u}r Quantenoptik und Quanteninformation, \"Osterreichische Akademie der Wissenschaften, 6020 Innsbruck, Austria
}
\affiliation{
    Institut f\"{u}r Experimentalphysik und Zentrum f\"{u}r Quantenoptik,\\ Universit\"{a}t Innsbruck, Technikerstra\ss e 25, 6020 Innsbruck, Austria
}
\author{Claudia Politi}
\affiliation{
    Institut f\"{u}r Quantenoptik und Quanteninformation, \"Osterreichische Akademie der Wissenschaften, 6020 Innsbruck, Austria
}
\author{Maximilian Sohmen}
\affiliation{
    Institut f\"{u}r Quantenoptik und Quanteninformation, \"Osterreichische Akademie der Wissenschaften, 6020 Innsbruck, Austria
}
\author{Philipp Ilzhöfer}
\affiliation{
    Institut f\"{u}r Quantenoptik und Quanteninformation, \"Osterreichische Akademie der Wissenschaften, 6020 Innsbruck, Austria
}
\author{Manfred J. Mark}
\affiliation{
    Institut f\"{u}r Quantenoptik und Quanteninformation, \"Osterreichische Akademie der Wissenschaften, 6020 Innsbruck, Austria
}
\affiliation{
    Institut f\"{u}r Experimentalphysik und Zentrum f\"{u}r Quantenoptik,\\ Universit\"{a}t Innsbruck, Technikerstra\ss e 25, 6020 Innsbruck, Austria
}
\author{Matthew A. Norcia}
\affiliation{
    Institut f\"{u}r Quantenoptik und Quanteninformation, \"Osterreichische Akademie der Wissenschaften, 6020 Innsbruck, Austria
}
\author{Francesca Ferlaino}
\affiliation{
    Institut f\"{u}r Quantenoptik und Quanteninformation, \"Osterreichische Akademie der Wissenschaften, 6020 Innsbruck, Austria
}
\affiliation{
    Institut f\"{u}r Experimentalphysik und Zentrum f\"{u}r Quantenoptik,\\ Universit\"{a}t Innsbruck, Technikerstra\ss e 25, 6020 Innsbruck, Austria
}

\begin{abstract}
    We report on the observation of heteronuclear magnetic Feshbach resonances in several isotope mixtures of the highly magnetic elements erbium and dysprosium. 
    Among many narrow features, we identify two resonances with a width greater than one Gauss.
    We characterize one of these resonances, in a mixture of \isotope[168]{Er} and \isotope[164]{Dy}, in terms of loss rates and elastic cross section, and observe a temperature dependence of the on-resonance loss rate suggestive of a universal scaling associated with broad resonances. Our observations hold promise for the use of such a resonance for tuning the interspecies scattering properties in a dipolar mixture.
    We further compare the prevalence of narrow resonances in an \isotope[166]{Er}-\isotope[164]{Dy} mixture to the single-species case, and observe an increased density of resonances in the mixture.
\end{abstract}

\maketitle



Ultracold quantum gases are a highly successful platform for physics research largely because it is possible to create simplified and controllable versions of condensed matter systems~\cite{Bloch:2008rev}. As the field has advanced, great progress has been made by reintroducing complexity in a carefully controlled manner. This complexity can manifest in the form of interparticle interactions~\cite{menotti:2008rev, Chin2010fri, saffman2010quantum}, the species and statistics of the particle under study~\cite{demarco1999onset, PhysRevLett.88.160401, taglieber2008quantum}, or in the form of the potential landscape, control protocols and imaging techniques applied to the system~\cite{Bakr2009qgm, gross2017quantum}. In this work, we explore interspecies Feshbach resonances as a means of generating tunable interactions between two different species of complex dipolar atoms.  



Atoms with large magnetic dipole moments, such as the lanthanide series elements erbium and dysprosium, interact in a manner that is both long-range and anisotropic. This is in contrast to more commonly used atomic species, such as alkali and alkaline earth metals, which primarily interact in a short-range and isotropic way. The recent creation of degenerate Bose and Fermi gases of such atoms~\cite{Lu:2011,Lu2012,Aikawa:2012,Aikawa2014} has enabled the observation of a wealth of new phenomena including quantum-stabilized droplet states~\cite{Kadau:2016,Chomaz:2016,Schmitt:2016}, roton quasi-particles~\cite{Chomaz:2018}, supersolid states~\cite{Bottcher2019,Chomaz:2019,Tanzi:2019}, and a non-isotropic Fermi surface~\cite{aikawa2014observation}.  

In a separate direction, degenerate mixtures of multiple atomic species have also provided diverse opportunities for the study of new physical phenomena. Examples include studies of polarons that arise when an impurity species interacts with a background gas~\cite{ospelkaus2006localization,will2011coherent, heinze2011multiband,spethmann2012dynamics,jorgensen2016observation, hu2016bose}, and the formation of heteronuclear molecules with large electric dipole moments~\cite{kohler:2006,ni2008high,takekoshi2014ultracold,molony2014creation}.  

We expect that combining dipolar interactions with heteronuclear mixtures will lead to a rich set of novel physical phenomena, the exploration of which has only recently begun. In particular, dipolar interactions are expected to have dramatic consequences for the miscibility of binary condensates~\cite{Gligoric2010, Wilson:2012, Kumar2017}, and in turn on vortex lattices that arise in such systems~\cite{Kumar2017a}. Further, novel properties of polarons are predicted to emerge when either the background~\cite{Kain:2014} or both background and impurity~\cite{Ardila2018} particles experience dipolar interactions~\cite{wenzel2018fermionic}. 

Dipolar heteronuclear mixtures have recently been demonstrated~\cite{Trautmann2018}, but so far the interspecies scattering properties have not been explored, either experimentally or theoretically. In these complex dipolar species, scattering properties are dictated by both anisotropic long-range dipolar interactions, which can be tuned through a combination of system geometry and magnetic field angle, and by contact interactions, which can be tuned through the use of interspecies Feshbach resonances. While scattering models and experimental demonstrations exist for mixtures of single- and two-valence electron atoms (which lack strong dipolar interactions)~\cite{stan2004observation, inouye2004observation}, the scenario of two multi-valence electron atoms has yet to be considered, and represents a new frontier for our understanding of ultracold scattering.  
In many commonly used atomic systems, the strength, character, and location of magnetic Feshbach resonances can be predicted with high precision through coupled-channel calculations~\cite{Chin2010fri}. However, the complexity of the internal level structure and coupling mechanisms present in lanthanide atoms lead to significant challenges for the development of a microscopic theory with predictive power, and so necessitate an experimental survey to find resonances with favorable properties~\cite{kotochigova2011anisotropy,Frisch:2014,baumann2014observation,Maier:2015,Khlebnikov:2019random}.  
 

To this end, we searched for heteronuclear Feshbach resonances broad enough to provide a practical means for tuning the interspecies interaction in Bose-Bose and Bose-Fermi dipolar quantum mixtures. Using atomic-loss spectroscopy to identify resonances, we perform surveys of fermionic \isotope[161]{Dy} and bosonic \isotope[164]{Dy} together with \isotope[166]{Er}, \isotope[168]{Er}, and \isotope[170]{Er} over a magnetic-field range from zero to several hundred Gauss (the exact range varies by isotope combination due to availability of favorable evaporation conditions). 
We also explored a Fermi-Fermi mixture of \isotope[167]{Er} and \isotope[161]{Dy}, but observed no broad resonances there. In Table~\ref{tab:survey} we summarize positions and widths of these features observed in our surveys. As an exemplary case, we present a more detailed characterization of the resonance near \unit 13.5 G in the \isotope[168]{Er}-\isotope[164]{Dy} Bose-Bose mixture, through measurements of interspecies thermalization and the dependence of atomic loss on temperature.

In addition, our dipolar mixtures host a large number of narrow interspecies resonances. In previous experiments with single species, the density and spacing of these narrow resonances has been studied to reveal a pseudo-random distribution that can be modeled well using random matrices~\cite{Frisch:2014,Maier:2015,Khlebnikov:2019random}.
By performing high resolution scans over specific magnetic-field ranges, we find that the average density of interspecies resonances exceeds the combined density of intraspecies resonances, perhaps indicating the contribution of odd partial waves or molecular states with antisymmetric electron configurations for the interspecies case, which are not present in the scattering of identical bosons.  

Finally, in each Fermi-Bose mixture involving \isotope[161]{Dy} we observe a correlated loss feature between fermionic Dy and bosonic Er atoms. Strangely, the loss feature is present at the same magnetic-field value for all three bosonic erbium isotopes studied. Such behavior is inconsistent with a typical interspecies Feshbach resonance, where the magnetic field at which the resonance occurs is strongly dependent on the reduced mass of the atoms involved~\cite{Frye:2019}. The mechanism behind this unusual feature is as of yet unknown and calls for further experimental and theoretical investigations.

\begin{table}[b]
\begin{ruledtabular}
\caption{
  Comparatively broad resonances found in specific isotope mixtures together with estimated center positions and widths (FWHM) from Gaussian fits to atom loss spectra. Each value is an average between the fit values of Er and Dy.
}
\label{tab:survey}
\begin{tabular}{c c c}
Combination & Resonance magnetic field (G) & Width (G) \\
\hline \\[-2ex]
\isotope[168]{Er}-\isotope[164]{Dy} & 13.32(4) &  1.7(1) \\ 
\isotope[166]{Er}-\isotope[164]{Dy} & 34.09(3) &  1.5(1) \\
\isotope[166]{Er}-\isotope[161]{Dy} & 161.31(3) &  0.84(9) \\ 
\isotope[168]{Er}-\isotope[161]{Dy} & 161.30(2) &  0.93(5) \\
\isotope[170]{Er}-\isotope[161]{Dy} & 161.26(3) &  0.91(8) \\
\end{tabular}
\end{ruledtabular}
\end{table}




Our experimental sequence is similar to the one introduced in our previous works~\cite{Ilzhoefer2018,Trautmann2018}. More details can be found in the supplemental material~\cite{supmat}. In brief, after cooling the desired isotope combination of erbium and dysprosium atoms into dual-species magneto-optical trap (MOT), we load the atoms into a crossed optical dipole trap (ODT) created by \unit 1064 nm laser light. Here we perform evaporative cooling down to the desired sample temperature. During the whole evaporation sequence, we apply a constant and homogeneous magnetic field ($B_{\rm{ev}}$), pointing along the $z$-direction opposite to gravity. $B_{\rm{ev}}$ preserves the spin-polarization into the lowest Zeeman sublevel of both species. We use different values of $B_{\rm{ev}}$ to optimize the evaporation efficiency depending on the isotope combination and on the range of the target magnetic field ($B_{\rm{FB}}$) to be investigated.
The final ODT has trap frequencies $\omega_{x, y, z} = 2 \pi \times (222,24,194)\,\si{\per\second}$. We typically obtain mixtures with atom numbers ranging from $\num{3e4}$ to $\num{1e5}$ atoms for each species. The sample is in thermal equilibrium at about \unit 500 nK , which corresponds to about twice the critical temperature for condensation. Typical densities are up to a few $\SI{e12}{cm^{-3}}$ for each species.   
After preparing the mixture, we linearly ramp the magnetic field from $B_{\rm{ev}}$ to $B_{\rm{FB}}$ in \unit 5 ms , either in an increasing or decreasing manner. We hold the mixture for a time ranging between \unit 5 ms and \unit 400 ms depending on the experiment. At the end of the hold time, we release the atoms from the ODT in a \unit 15 ms time-of-flight (TOF) expansion after which we record an image of the atoms using a standard low-field absorption imaging technique~\cite{supmat, Aikawa:2012}.
Note that we adjust the relative amount of erbium and dysprosium in the final thermal mixture for the specific experiments by independently tuning the MOT loading time for each species between \unit 0.5 s to \unit 5 s . 


\begin{figure}[ht]
    \centering
	\includegraphics[width=\columnwidth]{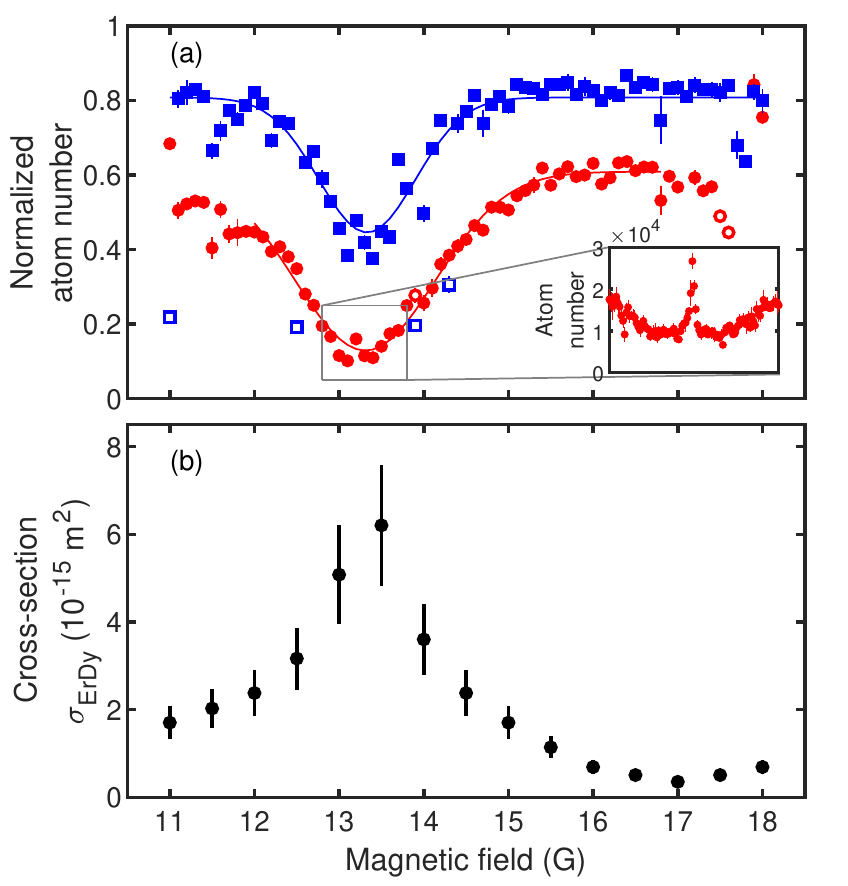}
	\caption {
    (a) Trap loss from the \unit 13.5 G resonance in the Bose-Bose mixture \isotope[168]{Er}-\isotope[164]{Dy} (red circles and blue squares points respectively). Empty symbols correspond to narrow single-species resonances, which we exclude from fits. Each point is an average over four experimental repetitions. For each magnetic field, the atom number recorded after \unit 200 ms of hold time is normalized to that at a short hold time of \unit 10 ms . The lines are the Gaussian fits to the data. The inset shows erbium loss measured in a different dataset with \unit 5 mG resolution, and highlights the structure present on the center of the feature. The same structure is visible also for the dysprosium atoms in the mixture. (b) Interspecies elastic cross-section $\sigma_{\rm{ErDy}}$ measured across the Feshbach resonance using cross-species thermalization. Each value of $\sigma_{\rm{ErDy}}$ is extracted from thermalization data using a numerical model for thermalization that includes temporal variation in atom number and temperature; see main text and supplemental material~\cite{supmat}.
	}
	 \label{fig:1} 
\end{figure}


In the isotope combinations and range of magnetic fields that we explore here, we observe two interspecies resonances with widths greater than \unit 1 G (see Table~\ref{tab:survey}). We now turn to a more detailed characterization of a feature present in the \isotope[168]{Er}-\isotope[164]{Dy} combination, for which atom loss is shown in \Myref[(a)]{fig:1}. We chose to focus on this feature because it is relatively isolated from the many narrow homonuclear and heteronuclear resonances typical of lanthanides. In this experiment, the starting mixture contains $\num{6.2e4}$ erbium and $\num{9.1e4}$ dysprosium atoms and it is prepared by evaporation at $B_{\rm{ev}}=\SI{10.9}{G}$. In order to compensate for loss during magnetic-field ramps and slow drifts of the atom number, we normalize measurements performed with \unit 200 ms hold times at $B_{\rm{FB}}$ to interleaved measurements at \unit 10 ms hold time at the same field. We further performed independent trap-loss spectra in single-species operation to confirm the interspecies nature of the resonance. Moreover, such scans allow us to identify intraspecies resonances and exclude them from the fit (see empty symbols in \Myref[(a)]{fig:1}). As shown in the inset for erbium, a high-resolution scan reveals a narrow region with less loss near the center of our broad loss feature, probably due to the influence of a second interspecies resonance. This structure is also visible on the dysprosium loss feature but it is not shown in the inset for ease of reading.

A Gaussian fit to the loss profiles, with known narrow single-species resonance excluded, returns a center value of \unit 13.31(2) G and \unit 13.33(4) G and a full width at half maximum value of \unit 1.95(5) G and \unit 1.3(1) G for erbium and dysprosium, respectively. The observed difference in the fitted width of the two species can be explained by the imbalance in atom number: because this measurement was performed with fewer erbium atoms than dysprosium, the fractional loss of erbium is higher than that of dysprosium, leading to a greater saturation of loss and broadening of the erbium loss feature.   


\begin{figure}[hb]
    \centering
	\includegraphics[width=\columnwidth]{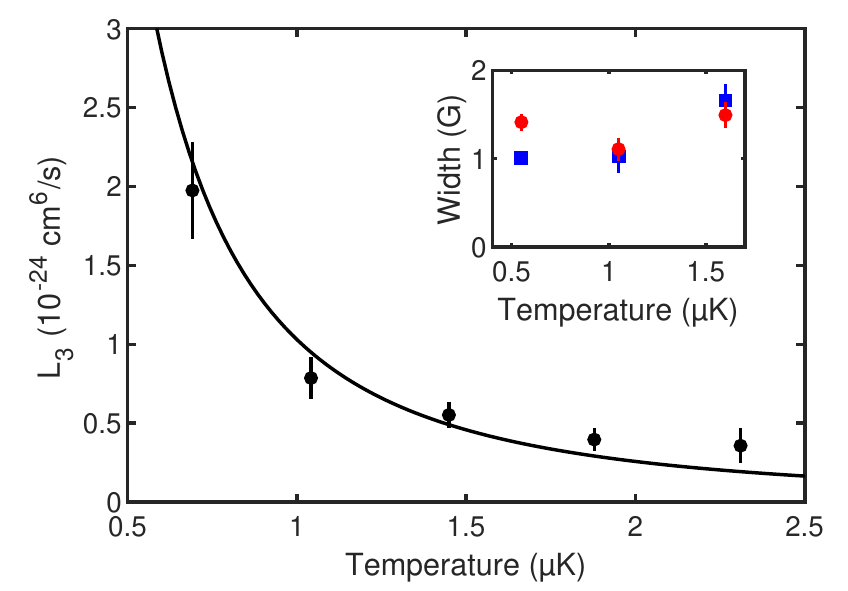}
	\caption {
    Three-body loss coefficient $L_3$ extracted from on-resonance loss measurements at the resonance position for different temperatures (black circles), along with a fit to a $1/T^2$ scaling (black line), as expected for universal three-body loss. The inset shows the resonance width extracted as FWHM from Gaussian fits to the trap-loss spectra versus cloud temperature for a different dataset. Red circles and blue squares refer to erbium and dysprosium respectively. The reported temperature comes from a TOF estimation.
	}
	 \label{fig:2} 
\end{figure}

To get insights on its effective strength and width, we perform cross-species thermalization measurements across the resonance (see \myref[(b)]{fig:1}). Interspecies thermalization experiments are well established techniques to extract effective thermalization cross sections, which in turn depends on the scattering length~\cite{Anderlini:2005,Guttridge:2017,Ravensbergen:2018}. While inferring on a precise value of the scattering length would require the development of a detailed and rigorous model that accurately captures the temperature-dependence of the interspecies and anisotropic dipolar scattering~\cite{bohn2014:therm}, and would go beyond the scope of this work, we are able to determine a thermally averaged scattering cross-section from which we can estimate the width of the resonance.

In this cross-thermalization experiment, we selectively heat dysprosium by means of a near-resonant \unit 421 nm light pulse along the vertical direction. We confirmed that the light pulse has no direct measurable effect on erbium. The magnetic field is then jumped to the desired value $B_{\rm{FB}}$ and held for a variable amount of time, during which the temperature of erbium rises to equilibrate with dysprosium due to elastic collisions. We record the temperature of the two species along a direction orthogonal to the heating pulse, as the effects of center of mass motion are less prevalent here~\cite{com:motion}, and use a numerical model to extract a cross section from the rate of thermalization~\cite{supmat}. This simple model assumes an energy independent cross section, an assumption which may break down near resonance where unitarity limits on scattering may become significant. 

From these thermalization measurements, we can see a dramatic increase in the scattering cross section near resonance, as one would expect for an interspecies Feshbach resonance. Further, we observe a significant modification of the cross section associated with the resonance over a Gauss-scale range of magnetic fields, similar to the width we observe in loss measurements. For an isolated resonance and pure contact interactions, a common way to characterize the resonance width is the parameter $\Delta$, given by the difference in magnetic field between the pole of the resonance, at which the thermalization rate is maximal, and the nearest zero-crossing in the thermalization rate, which would correspond to a lack of scattering~\cite{Chin2010fri}. In lanthanides, the presence of anisotropic dipolar interactions leads to a scattering cross section that does not completely vanish. In addition, multiple narrow and overlapping resonances may be present, which may influence the interpretation of such a width measurement. However, to get a rough estimate of the width of the resonance, we can consider the distance between the resonance pole and the apparent minimum in the thermalization rate at \unit 17 G . This suggests a width of $\Delta \simeq \SI{3.5}{G}$.  

The dependence of the loss feature on the cloud temperature can provide additional information on the nature of the resonance. For broad resonances, a universal regime is expected to emerge near resonance where the scattering cross section and loss are dictated primarily by the atomic momentum, rather than the scattering length~\cite{rem2013lifetime}. In this regime, the three-body loss parameter $L_3$ follows a nearly universal form scaling as $1/T^2$, where $T$ is the temperature. Such scaling has been observed in broad resonances of several atomic species~\cite{rem2013lifetime,maier2015broad,eismann2016universal}.  

We observe a temperature dependence of the loss rate near resonance that is suggestive of such universal behavior. By varying the final depth of the ODT reached during evaporation, we tune the temperature of the atomic mixture. For each temperature, we measure atom loss on resonance at \unit 13.4 G as a function of the hold time. We then use a numerical model to extract the rate of interspecies three-body loss, and $L_3$~\cite{supmat}. 

These loss coefficients are plotted as a function of temperature in \myref[]{fig:2}, along with a fit to a $1/T^2$ dependence, which provides a reasonable description of our data. 
The universal temperature dependence arises from a maximum value of $L_3$ possible at a given temperature, given by:
\begin{equation}
L_{3,\rm{max}} = \frac{\lambda_{3,\rm{max}}}{T^2} \simeq \frac{\hbar^5}{m^3}\frac{36\sqrt{3}\pi^2}{(k_{\rm{B}} T)^2} .
\end{equation}
Factors associated with Efimov physics~\cite{supmat} can lead to a lower value for $L_3$, but not higher~\cite{helfrich2010three,rem2013lifetime, d2018few}. From our fit to a $1/T^2$ dependence for our data, we extract a value of $\lambda_3 = \SI{1.0(2)e-24}{\micro K^2 cm^6 s^{-1}}$, which is compatible with the predicted bound of $\lambda_{3,\rm{max}} = \SI{2.4e-24}{\micro K^2 cm^6 s^{-1}}$.

A reduction in the peak loss rate with increasing temperature can also result from thermal broadening of the resonance, especially for very narrow resonances~\cite{Maier:2015}. This is unlikely to be the dominant effect here, as for typical differential magnetic moments between entrance and closed channels in our lanthanide system~\cite{Frisch:2015mol}, we would expect broadening on the scale of a few times \unit 10 mG for temperatures near \unit 1 {\micro K} , much narrower than the Gauss-scale width of our feature. Further, suppression of peak loss is typically accompanied by a commensurate broadening and shift of the loss feature on the scale of its width, which we do not observe (inset in \myref[]{fig:2}).  


\begin{figure*}[ht]
    \centering
	\includegraphics[width=\textwidth]{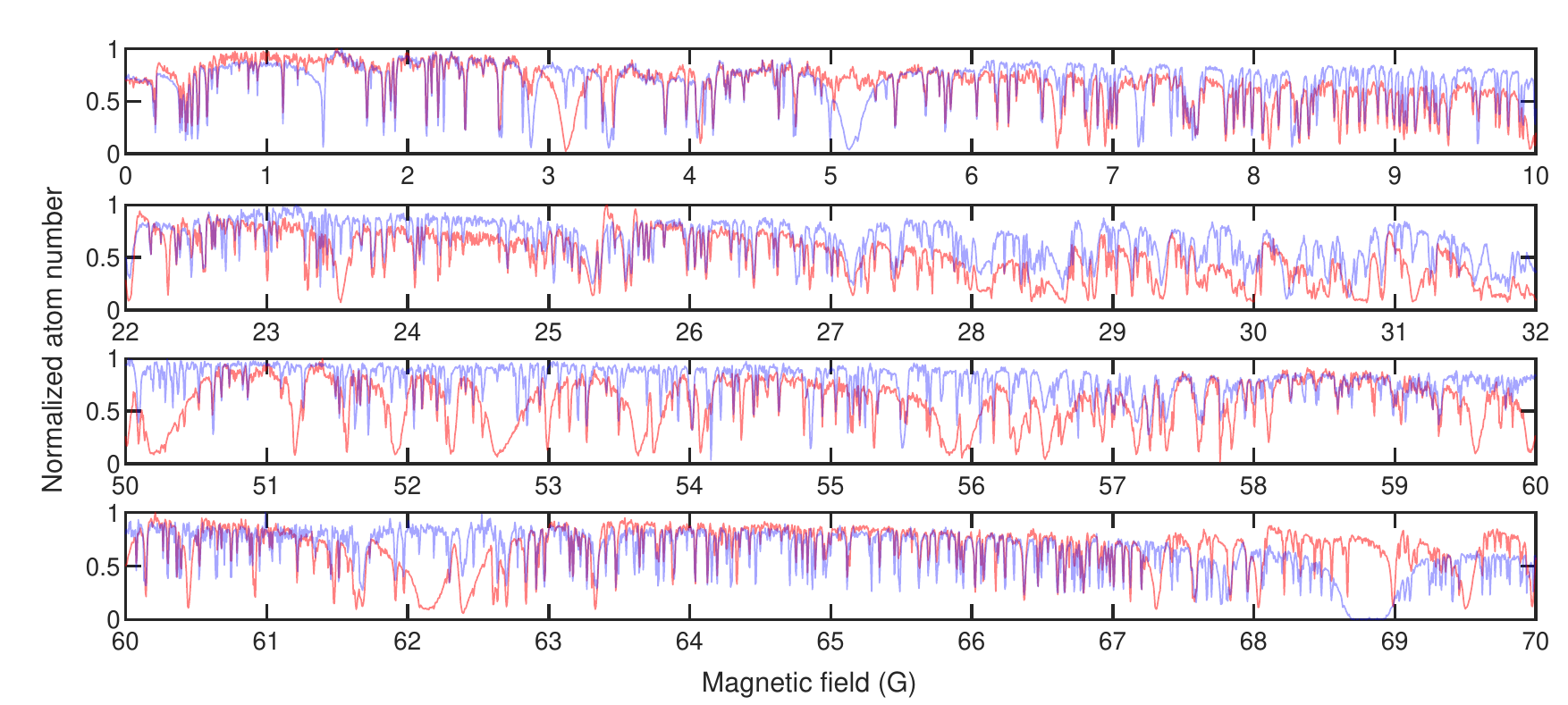}
	\caption {
	High-resolution trap-loss spectroscopy for a balanced mixture of \isotope[166]{Er} and \isotope[164]{Dy} (red and blue curves respectively), with initial atom numbers of roughly $10^5$ per species and a temperature of \unit 500 nK and after \unit 400 ms of interaction time. The measurement is composed of four datasets \range 0 10 G , \range 22 32 G , \range 50 60 G , and \range 60 70 G with a stepsize of \unit 5 mG. Each point is an average over four experimental repetitions. Atom numbers are normalized to the maximum of each dataset for ease of reading. The broad loss feature in Dy near \unit 68.8 G was not observed in previous work~\cite{Maier:2015}, and may result from a technical source of loss in our experiment.
	}
	 \label{fig:3} 
\end{figure*}

In addition to the few relatively broad resonances, the lanthanides exhibit many narrow resonances, whose statistical properties have been investigated for single-species gases~\cite{Frisch:2014, Maier:2015,Khlebnikov:2019random}. In this section we compare the abundance of interspecies resonances to single-species resonances by performing high-resolution trap-loss spectroscopy on the isotope combination \isotope[166]{Er}-\isotope[164]{Dy} (see \myref[]{fig:3}). Here, we investigate four different magnetic-field ranges, each \unit 10 G wide,
with a resolution 40 times higher than the one used for the exploratory surveys. 
To enable direct comparison with the previous works performed on single species~\cite{Frisch:2014,Maier:2015}, we use similar experimental conditions (isotope, atom number, temperature, and hold time). 



\begin{figure}[ht]
   \centering
	\includegraphics[width=\columnwidth]{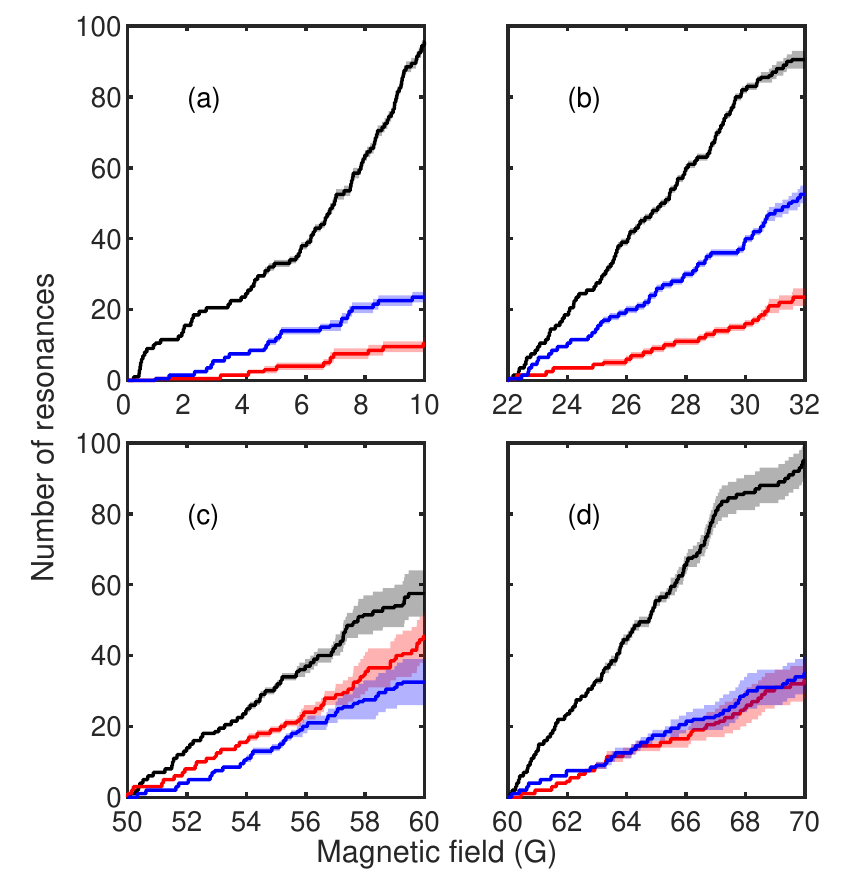}
	\caption {
	(a-d) Staircase function describing the number of Feshbach resonances as a function of the four investigated magnetic-field ranges: \range 0 10 G , \range 22 32 G , \range 50 60 G , and \range 60 70 G respectively. The black line shows the number of heteronuclear resonances. The red and blue lines show the number of homonuclear resonances for \isotope[166]{Er} and \isotope[164]{Dy}, respectively. The shaded areas represent our confidence intervals (see main text).
	}
	 \label{fig:4} 
\end{figure}


As expected, we observe many narrow homonuclear resonances~\cite{Frisch:2014, Maier:2015}. In addition, we also identify many narrow heteronuclear resonances. To distinguish these two types of resonance, we first label features with a fractional loss above 30\% as resonances. We then categorize these resonances as interspecies if erbium and dysprosium loss features occur simultaneously within a range of $\SI{\pm 10}{mG}$ and with a loss amplitude ratio in the range 0.5-2. Features that do not meet both of these criteria, are labelled either as homonuclear or ambiguous, based on comparison with separate scans performed with single species, either within this work or from previously published data~\cite{Frisch:2014, Maier:2015}. The number of ambiguous features define our confidence interval.


In order to visualize the number of resonances, we construct the staircase function $\mathcal{N}(B)$, which describes the cumulative number of resonances from the start of a scan range up to a given magnetic field $B_{\rm{FB}}$. \Myref[(a-d)]{fig:4} shows $\mathcal{N}(B)$ for the four investigated magnetic-field ranges. The black lines represent heteronuclear Feshbach resonances, while the blue and the red lines represent the homonuclear \isotope[166]{Er} and \isotope[164]{Dy} resonances, respectively. The shaded regions represent our confidence interval defined by the total number of ambiguous Feshbach resonances. 

Our analysis results in a total number of heteronuclear resonances of $\mathcal{N}_{\rm{ErDy}}(\rm{tot})=339(16)$, counting all magnetic-field ranges, and a number of homonuclear resonances of $\mathcal{N}_{\rm{Er}}(\rm{tot})=116(16)$ and $\mathcal{N}_{\rm{Dy}}(\rm{tot})=144(16)$.
Within our confidence intervals, we detect a total number of homonuclear resonances comparable with those of previous works~\cite{Frisch:2014,Maier:2015}.  
The corresponding total density of resonances $\Bar{\rho}$, given by the total number of resonances divided by the total range of magnetic fields scanned are: $\Bar{\rho}_{\rm{ErDy}} = \SI{8.5(4)}{G^{-1}}$, $\Bar{\rho}_{\rm{Er}}=\SI{2.9(4)}{G^{-1}}$, and $\Bar{\rho}_{\rm{Dy}}=\SI{3.6(4)}{G^{-1}}$.


For our combined dataset, we find that the total number of heteronuclear resonances exceeds the combined number of homonuclear resonances for the two species: $\Bar{\rho}_{\rm{ErDy}} = \alpha (\Bar{\rho}_{\rm{Er}} + \Bar{\rho}_{\rm{Dy}})$, with $\alpha = 1.3(2)$. We would expect that the average density of heteronuclear resonances should be greater than the sum of the two homonuclear resonance densities. This is because each species contributes a set of internal states that can be coupled to, and the heteronuclear resonances are not subject to the same symmetrization requirements as the homonuclear resonances. 
In resonances involving distinguishable particles, both gerade and ungerade Born-Oppenheimer molecular potentials contribute, as well as both even and odd partial waves for the entrance channel. Our data is consistent with this expectation ($\alpha > 1$). Note that we do observe a lower number of interspecies resonances in the range \range 50 60 G , perhaps as a result of the non-random distribution of resonances as observed in the single-species case~\cite{Frisch:2014,Maier:2015}, or to the presence of broad homonuclear erbium resonances that could obscure the observation of interspecies resonances.




\begin{figure}[ht]
    \centering
	\includegraphics[width=\columnwidth]{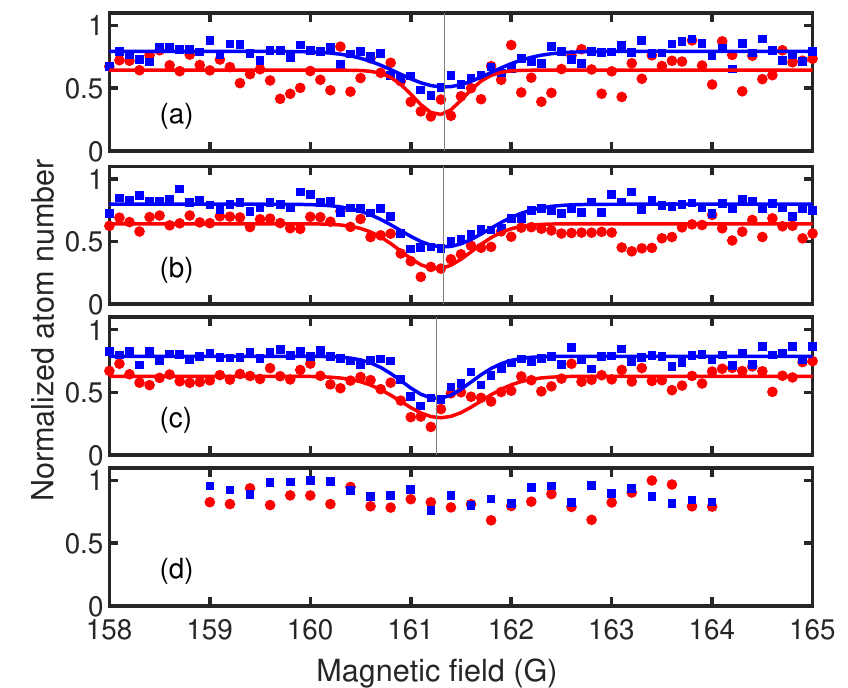}
	\caption {
	Trap loss spectra for fermionic \isotope[161]{Dy} in combination with bosonic \isotope[166]{Er}, \isotope[168]{Er}, and \isotope[170]{Er}, and fermionic \isotope[167]{Er} (a-d, respectively). Red circles represent erbium, blues squares represent dysprosium and lines are Gaussian fits to the losses. The solid vertical gray lines highlight the peak centers from the fit over dysprosium losses. For the plots with bosonic erbium, the atom number after \unit 100 ms of interaction time is normalized to a short hold time of \unit 5 ms . In the plot with \isotope[167]{Er}, the normalization is performed using the maximum value in the dataset. For all panels, each point is an average over four experimental repetitions.
	}
	 \label{fig:5} 
\end{figure}

Finally, we have also searched for broad (Gauss-range) resonances in Bose-Fermi mixtures consisting of fermionic \isotope[161]{Dy} combined with different bosonic isotopes of erbium -- \isotope[166]{Er}, \isotope[168]{Er}, and \isotope[170]{Er}, as well as Fermi-Fermi mixtures of \isotope[161]{Dy} and \isotope[167]{Er}. For these combinations, we perform only coarse scans and thus only resolve broad features. In mixtures involving the bosonic isotopes of erbium we observe a correlated loss feature between erbium and dysprosium near \unit 161 G (see \myref[]{fig:5}). This loss feature is not present at our level of measurement sensitivity with either species alone, or in the mixture with the fermionic \isotope[167]{Er}. Surprisingly, the loss feature is centered at the same magnetic field (to within our resolution of \unit 0.1 G ) for all bosonic isotopes of erbium. This is quite unexpected as the magnetic-field value of the resonance position is typically highly sensitive to the reduced mass of the atoms involved~\cite{Frye:2019}.  


Several physical mechanisms could be consistent with such a feature. One possibility is that the resonance we observe is associated with a bound state of a shallow molecular potential~\cite{pc}. Mechanisms to create such potentials have been proposed for species with dipolar interactions~\cite{avdeenkov2004field, karman2018near}. However, none are obviously applicable to magnetic atoms in the lowest energy entrance channel. Further, given the level of insensitivity to the mass of erbium, we would expect to see additional resonances of a shallow potential in the magnetic-field range over which we survey, which we do not. A second possibility is that the feature we observe is not a true interspecies resonance, but rather an intraspecies resonance in dysprosium whose loss rate is enhanced by the presence of bosonic erbium atoms. A similar effect was reported in a mixture of fermionic lithium and bosonic rubidium atoms~\cite{deh2008feshbach}. Finally, it is possible that this feature is not a Feshbach resonance at all, but rather the result of spin-changing processes resulting from unintentional radio-frequency tones in the laboratory, or of an interspecies photoassociation resonance. We have ruled out the most likely culprits for the last effect by varying the relative detuning between our horizontal and vertical dipole traps and observing no change in the resonance position. We hope that our presentation of this mysterious feature may spur theoretical exploration of possible physical mechanisms.  


In conclusion, we have reported experimental observation of heteronuclear magnetic Feshbach resonances in several isotope mixtures of erbium and dysprosium. Among the Gauss-broad features identified in our surveys, we have characterized one in the combination \isotope[168]{Er}-\isotope[164]{Dy} by means of cross-species thermalization measurement and temperature dependence analysis. We performed high-resolution trap-loss spectroscopy in the combination \isotope[166]{Er}-\isotope[164]{Dy} to compare the average resonance density of the mixture with respect to the single-species case. In mixtures of fermionic \isotope[161]{Dy} and bosonic erbium atoms, we observed a correlated loss feature which appears to be insensitive on the erbium isotope used but absent in dysprosium alone. Our observations pave the way to realize tunable interactions in quantum degenerate mixtures of dipolar atoms, which will enable varied opportunities including studies of the miscibility of binary condensates, of vortex lattices, and of dipolar polarons~\cite{Gligoric2010,Wilson:2012,Kain:2014,Kumar2017,Kumar2017a,Ardila2018}.

\begin{acknowledgments}
 
We thank Jeremy Hutson, Matthew Frye, John Bohn, Arno Trautmann, and the Erbium and DyK teams in Innsbruck for insightful discussions. This work is financially supported through an ERC Consolidator Grant (RARE, No.\,681432), a NFRI Grant (MIRARE, No.\,\"OAW0600) from the Austrian Academy of Science, a QuantERA grant MAQS by the Austrian Science Fund FWF No\,I4391-N, and a DFG/FWF (FOR 2247/PI2790). M.~S.~and G.~D.~acknowledge support by the Austrian Science Fund FWF within the DK-ALM (No.\,W1259-N27). We also acknowledge the Innsbruck Laser Core Facility, financed by the Austrian Federal Ministry of Science, Research and Economy.

\end{acknowledgments}

* Correspondence and requests for materials should be addressed to Francesca.Ferlaino@uibk.ac.at.

\bibliography{BibFeshbach}


\clearpage
\appendix
\renewcommand\thefigure{\thesection S\arabic{figure}}   
\setcounter{figure}{0}   

\section{Supplemental Material}

\subsection{Preparation}
\label{sup:prep}

The experimental sequence is similar to the one introduced in our previous works~\cite{Ilzhoefer2018,Trautmann2018}. After cooling the erbium and dysprosium atoms into dual-species magneto-optical traps (MOTs) of the desired isotope combination, we load about $1-5 \times 10^{6}$ atoms of both erbium and dysprosium into a single-beam optical dipole trap created by \unit 1064 nm laser light and horizontally propagating along the $y$-direction (hODT). We perform an initial stage of evaporative cooling of about \unit 0.8 s . After that, a second trap beam, coming from the same laser source but detuned by \unit 220 MHz , is shone along the vertical direction $z$ (vODT) onto the atoms forming a crossed ODT where we continue the evaporation for an additional duration of about \unit 4.3 s down to the desired sample temperature. During the whole evaporation sequence, a constant and homogeneous magnetic field ($B_{\rm{ev}}$) pointing along the $z$-direction and opposite to gravity is applied. Different values of $B_{\rm{ev}}$ are used depending on the isotope combination and on the range of the target field ($B_{\rm{FB}}$) to be investigated. We typically end up with $3-10\times 10^{4}$ atoms for each species, in thermal equilibrium at about \unit 500 nK (about twice the critical temperature for condensation). Final trap frequencies are $\omega_{x, y, z} = 2 \pi \times (222,24,194)\,\si{\per\second}$.

At this point, we linearly ramp the magnetic field from $B_{\rm{ev}}$ to the target field $B_{\rm{FB}}$ in \unit 5 ms , either in an increasing or decreasing manner. We hold the mixture for a specific time ranging between \unit 5 ms and \unit 400 ms depending on the experiment. At the end of the hold time, we release the atoms from the ODT in a \unit 15 ms time-of-flight (TOF) expansion after which we record an image of the atoms using a standard low-field absorption imaging technique~\cite{Aikawa:2012}. We use pulses of resonant light in the horizontal $x-y$ plane at an angle of $\sim 45\,^\circ$ with respect to our weak trap axis $y$. \unit 5 ms after the clouds being released from the ODT, we linearly ramp $B_{\rm{FB}}$ to zero in \unit 10 ms . At the same time, we activate a magnetic field of about \unit 4 G pointing along the imaging direction.

Note that we select the relative amount of erbium and dysprosium required in the final thermal cloud by the specific experiments by independent tuning of the respective MOT loading times between \unit 0.5 s to \unit 5 s . $B_{\rm{ev}}$ and $B_{\rm{FB}}$ are generated by the same pair of coils in Helmholtz configuration. We ramp them up linearly in the early stage of the evaporation sequence \unit 200 ms after loading the atoms into the hODT. We checked that the response of the current flowing in the coils can follow time ramps on the millisecond time-scale. This translates in a effective change of the field at the sample position in the order of \unit 10 ms to settle at the part per thousand level.

\subsection{Cross-species thermalization}

\begin{figure}[ht]
   \centering
	\includegraphics[width=\columnwidth]{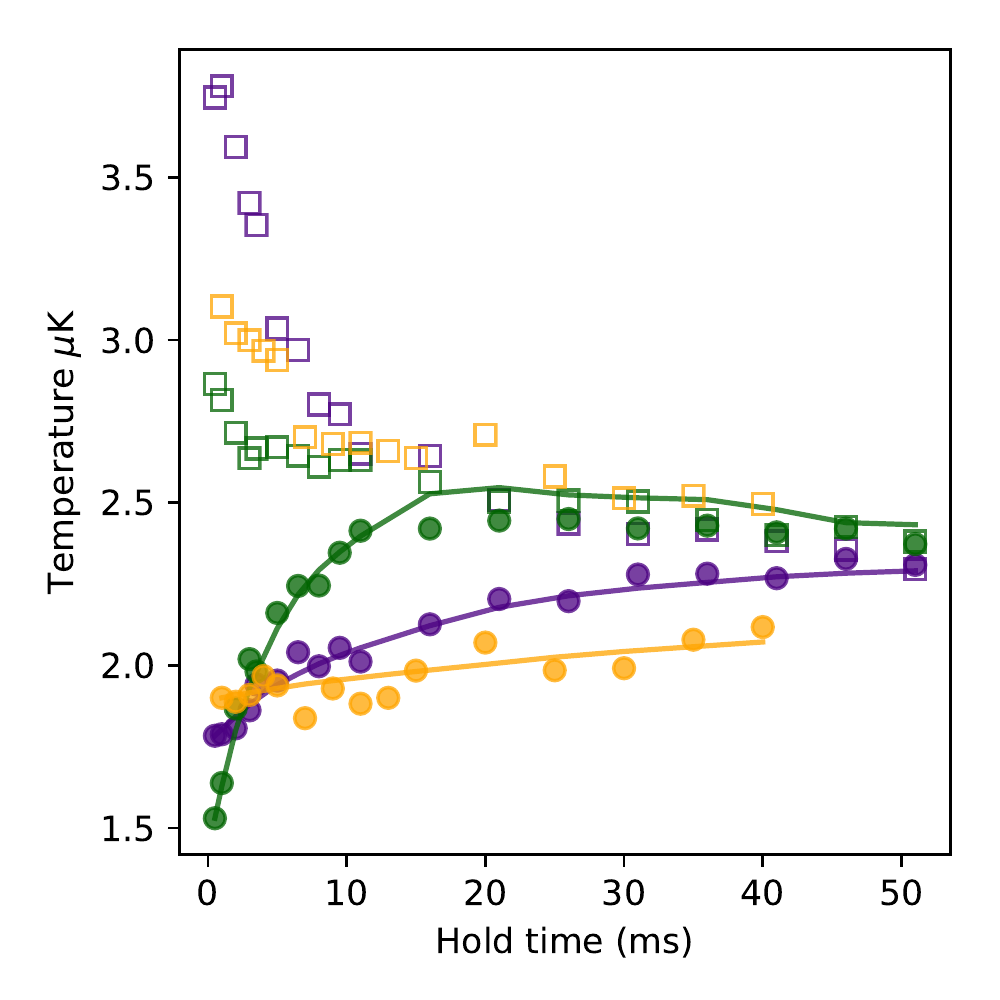}
	\caption {Sample temperature traces for erbium (filled circles) and dysprosium (hollow squares) after dysprosium is heated. Purple, green, and orange correspond to magnetic fields of \unit 12 G , \unit 13.5 G , and \unit 17 G , respectively. Fit lines represent the results of the numerical integration of equation~\ref{eq:Xtherm}, which fits the temperature profile of erbium based on its initial value and the dysprosium temperatures. Different evaporation conditions cause the curves to have slightly different initial and final conditions (see main text).
	}
	 \label{fig:Xtherm} 
\end{figure}

As an exemplary case, we study in more detail the resonance found in the \isotope[168]{Er}-\isotope[164]{Dy} Bose-Bose mixture, near \unit 13.5 G . To quantify reliably the value of the interspecies cross-section, we developed the following scheme for cross-species thermalization measurements~\cite{Anderlini:2005,Guttridge:2017,Ravensbergen:2018}. To avoid heating of the sample by crossing Feshbach resonances, we evaporate the mixture at $B_{\rm{ev}}$ close to resonance. Specifically, when measuring on the low(high)-field side of the feature we evaporate at $B_{\rm{evap}} = \SI{10.8}{G} (\SI{16.4}{G})$. Once the sample is prepared as previously described (here we use an unbalanced mixture with twice as much Dy as Er), we compress the trap by linearly increasing the hODT power by a factor of five and the vODT power by two in \unit 500 ms to prevent any plain evaporation. The final trap frequencies in the compressed trap are $\omega_{x, y, z} = 2 \pi \times (409,26,391)\,\si{\per\second}$. Subsequently, we ramp the magnetic field in \unit 5 ms to either \unit 10 G or \unit 16 G . Here, a pulse of near-resonant \unit 421 nm light propagating along the magnetic field direction ($z$) is used to selectively heat dysprosium. We fix the duration of the pulse at \unit 5.5 ms to roughly match the trap oscillation period along this direction and set the pulse intensity to give the desired temperature increase of the dysprosium cloud (up to \unit 4 {\micro K} ). We confirmed that the light pulse has no direct measurable effect on erbium. Finally with a quench fast compared to the shortest thermalization rate, the magnetic field is set to the desired value $B_{\rm{FB}}$ and held for a variable amount of time, during which the temperature of erbium rises to equilibrate with dysprosium due to thermalizing collisions (\Myref[]{fig:Xtherm}). We note that in the temperature evolution of the clouds, the initial temperatures are slightly different. This behavior is mainly due to different evaporation conditions on the two sides of the resonance, the different strength in the quench to the final $B_{\rm{FB}}$, and the heating caused by the resonance itself. By comparing the two species' temperature, we ensure that these different conditions are consistent with general offsets on the single measurement thus not affecting the final estimation of the cross-section.

To extract a scattering cross-section from our cross-species thermalization data, we use a fit to a numerical model for the thermalization of two species. In principle, a simple exponential fit to the temperature difference between the two species could also be used, but does not account for changes in the atom number or average temperature of the sample that may arise from residual evaporation during the thermalization time. Our numerical model follows that of Ref.~\cite{Anderlini:2005}. We treat the scattering cross section as independent of the energy of the colliding particles, an assumption that greatly simplifies the analysis, but inevitably breaks down near enough to resonance where unitarity considerations bound the scattering cross section. This assumption leads to a collision rate for each atom of species 1 with atoms of species 2 given by:
\begin{equation}
    \gamma_{12} = \frac{N_2 m_1^{3/2} \bar{\omega}_1^3}{\pi^2 k_{\rm{B}} (T_1 + \beta^{-2} T_2)^{3/2}}\sqrt{\frac{T_1}{m_1} + \frac{T_2}{m_2}}\sigma_{12}
\end{equation}
where $m_1$, $m_2$, $T_1$ and $T_2$ are the masses and temperatures of species 1 and 2, $\bar{\omega} = (\omega_x \omega_y \omega_z)^{1/3}$ characterizes the frequency of the trap, $\beta^2 = m_2 \bar{\omega}_2^2/m_1 \bar{\omega}_1^2$, and $\sigma_{12}$ is the effective interspecies cross section. We assume that the energy exchanged per collision is given by $\Delta E = \xi k_{\rm{B}} (T_2-T_1)$ where $\xi = 4 m_1 m_2/(m_1 + m_2)^2$, and that the heat capacity of each atom is $3 k_{\rm{B}}$. This leads to a differential equation for the temperature of erbium:
\begin{equation}
\begin{split}
    \frac{d T_{\rm{Er}}}{dt} = \frac{\xi k_{\rm{B}}(T_{\rm{Dy}} -T_{\rm{Er}})N_{\rm{Dy}} m_{\rm{Er}}^{3/2} \bar{\omega}_{\rm{Er}}^3}{3\pi^2 k_{\rm{B}}(T_{\rm{Er}} + \beta^{-2} T_{\rm{Dy}})^{3/2}} \\
    \times \sqrt{\frac{T_{\rm{Er}}}{m_{\rm{Er}}} + \frac{T_{\rm{Dy}}}{m_{\rm{Dy}}}}\sigma_{\rm{ErDy}}
\end{split}
\label{eq:Xtherm} 
\end{equation}
which we can numerically integrate using the instantaneous values for $T_{\rm{Dy}}$ and $N_{\rm{Dy}}$, and from this extract the scattering cross section $\sigma_{\rm{ErDy}}$ that yields a thermalization profile that best matches our data, as determined through a least-squares difference. Examples of three such fits, for \unit 12 G , \unit 13.5 G , and \unit 17 G are shown in \myref[]{fig:Xtherm}, and generally describe our thermalization data well.  

\subsection{Temperature dependence of loss}

We quantify the temperature dependence of three-body loss in terms of the interspecies three-body loss coefficient. For a single species, the three-body loss coefficient $L_3$ can be defined by: $\dot{N}/N = -L_3 \langle n^2 \rangle$ where $N$ is the total number of atoms, and $\langle n^2 \rangle = \int d^3r \ n^3(\boldsymbol{r})/N$ represents the average squared density of scattering partners for an atom in the gas. $n(\boldsymbol{r})$ is the local density of the gas.  

We define analogous quantities for our two-species mixture, containing particles denoted $i$ and $j$. In this case, 
\begin{equation}
\label{eq:Nloss}
\begin{split}
\frac{\dot{N}_i}{N_i} = \frac{-1}{3N_i} \int d^3r \ (2L_3^{i,i,j} n_i^2(\boldsymbol{r})n_j(\boldsymbol{r}) \\
+ L_3^{j,j,i} n_i(\boldsymbol{r})n_j^2(\boldsymbol{r}) ) .
\end{split}
\end{equation}
Here, $L_3^{i,i,j}$ represents the loss rate due to collisions involving two atoms of species $i$ and one of $j$.

To arrive at simple expressions, we make several assumptions and approximations. First, we treat the mass, temperature, and polarizability of the two atomic species as equal, which is a reasonable approximation for erbium and dysprosium isotopes in our \unit 1064 nm wavelength ODT~\cite{Trautmann2018}. This assumption implies equivalent spatial distributions for the two species, which we assume to be thermal in our three-dimensional harmonic trap. We next set $L_3^{i,i,j} = L_3^{j,j,i} \equiv L_3^i$ near resonance, essentially assuming that the loss process is primarily determined by the two pairwise interactions between the minority participant and the two majority atoms. We find this assumption leads to a model consistent with our observed relative loss between the two species. With these simplifications in place, we define $L_3^i$ using: $ \dot{N}_i/N_i = - L_3^i \langle n^2 \rangle_{\rm{eff}}^i $, where
\begin{equation}
     \langle n^2 \rangle_{\rm{eff}}^i = \frac{(2 N_i N_j + N_j^2)m^3 \bar{\omega}^6}{3^\frac{5}{2} 8 \pi^3 (k_{\rm{B}} T)^3}
\end{equation}
and $\bar{\omega} = (\omega_x \omega_y \omega_z)^{1/3}$ is the geometric mean of the trap oscillation frequencies.  

We extract the resonant value of $L_3$ by measuring remaining atom number versus hold time in mixtures prepared at different temperatures, with the magnetic field set near resonance at \unit 13.4 G . We then fit the resulting data by numerically integrating Eq.~\ref{eq:Nloss}. Because we observe significant single-species loss of erbium (the majority species), we treat the erbium atom number measured at each time-step as inputs to our fit, and extract the value of $L_3$ that best predicts the loss of dysprosium. Here, we assume that $L_3^{i,i,j} = L_3^{j,j,i} \equiv L_3$. We bound the effects of single-species loss in dysprosium by repeating the same measurement and analysis protocol off resonance at \unit 11.5 G and \unit 16.5 G . The error bars in \myref[]{fig:2} of the main text include a contribution corresponding to the extracted $L_3$ in the off-resonant condition, which contain both the effects of single-species loss and the small effect of off-resonant interspecies loss. Also included are errors associated with the observed change in temperature during the loss measurement, and relating to the approximations made in estimating the density. 


In a regime where the scattering length $a$ exceeds the thermal wavelength $\lambda_{\rm{th}} = h/\sqrt{2 \pi m k_{\rm{B}} T}$, and thermal broadening is small compared to the width of the loss feature, we expect roughly $L_3 \propto 1/T^2$, as has been observed in several experiments involving single atomic species~\cite{rem2013lifetime,maier2015broad,eismann2016universal}. This picture becomes complicated somewhat in the case of a binary mixture due to stronger Efimov effects, which lead to a temperature-dependent modulation of loss relative to the simple $1/T^2$ prediction. In particular, the parameter $s_0$, which characterizes the strength of the three-body Efimov potential, is equal to approximately 1.006 for identical bosons, but approximately 0.41 for our binary mixture~\cite{d2018few,helfrich2010three}. The fractional importance of these temperature-dependent modifications scale as $e^{-\pi s_0}$~\cite{rem2013lifetime}, making them a minor correction for identical bosons, but a potentially important effect in mixtures. It is possible that such effects contribute to deviations of our data from a $1/T^2$ form, but a true calculation would require knowledge of short-range inelastic processes in our system.



\end{document}